\documentclass[preprint,showkeys, preprintnumbers,amsmath,amssymb, graphicx,  graphics, epsfig]{revtex4}
\input{epsf}
\usepackage{graphicx}
\usepackage{amssymb}
\usepackage{bm}

\begin{document}

\def\be{\begin{equation}}
\def\ee{\end{equation}}
\def\bdm{\begin{displaymath}}
\def\edm{\end{displaymath}}
\def\erfc{\hbox{erfc }}
\def\vpa{v_{\parallel }}
\def\vper{v_{\perp }}
\def\Omm{\Omega }
\def\ppa{p_{\parallel }}
\def\pper{p_{\perp }}
\def\ppv{\vec{p}}
\def\kkv{\vec{k}}
\def\omm{\omega }

\title{\bf Surface waves on a quantum plasma half-space}

\author{M. Lazar$^{1,2}$ \footnote{mlazar@tp4.rub.de} P.K. Shukla$^{2,3}$ and A. Smolyakov$^1$} 
\affiliation{$^1$ Department of Physics and Engineering Physics, University of Saskatchewan, 116 Science Place, 
Saskatoon, Saskatchewan S7N 5E2, Canada 
\\ $^2$ Institut f\"ur Theoretische Physik, Lehrstuhl IV:
Weltraum- und Astrophysik, Ruhr-Universit\"at Bochum, D-44780 Bochum, Germany
\\$^3$ School of Physics, University of KwaZulu - Natal, Durban 4000, 
South Africa}

\date{\today}
\begin{abstract}

Surface modes are coupled electromagnetic/electrostatic 
excitations of free electrons near the vacuum-plasma interface
and can be excited on a sufficiently dense plasma half-space.
They propagate along the surface plane and decay in either 
sides of the boundary. In such dense plasma models, which are 
of interest in electronic signal transmission or in some 
astrophysical applications,
the dynamics of the electrons is certainly affected by the quantum effects.
Thus, the dispersion relation for the surface wave on a quantum electron plasma 
half-space is derived by employing the quantum hydrodynamical (QHD) and 
Maxwell-Poison equations. 
The QHD include quantum forces involving the Fermi electron temperature and
the quantum Bohm potential. It is found that, at room temperature, the 
quantum effects are mainly relevant for the electrostatic surface 
plasma waves in a dense gold metallic plasma.
\end{abstract}

\keywords{quantum plasma -- surface plasma waves (surface plasmon polariton) -- electrostatic surface waves (surface plasmon)}
\maketitle

\section{Introduction}

In the last years, there has been tremendous progress in nanotechnology,
mainly oriented towards the design of the novel quantum electronic devices \cite{yv06, cs06, sk07}. 

Light waves would transmit data at frequencies 10$^5$ times faster than 
today's Pentium chips, but using photons to carry data across a computer 
chip is currently impossible due to the optical diffraction in metals or semiconductors, which 
obstructs the propagation of light ($\lambda > 500$ nm) in 
electronic circuits miniaturized at dimensions below 100 nm.
However, using plasmons one can to make devices 
the same size as electrical components but give them the speed of photons \cite{ld02, ma05}. 
Due to difficulties in fabrication of three-dimensional transmission
structures in the high frequency (optical) spectral range, two-dimensional
structures are mainly being considered \cite{ma05}. Therefore, alternative
approaches are based on the physics of the so-called surface plasmon polariton \cite{b74}.
A surface plasmon polariton (SPP) is excited at frequencies lower than the plasma
frequency ($\omega < \omega_{pe}$), as a hybrid surface mode 
resulting from the linear coupling of its 
electromagnetic and electrostatic components: the electromagnetic wave is trapped on 
the surface because of the interaction with the free electrons of the plasma. 
Concluding, we can define a SPP as a combined excitation
consisting of a surface plasmon and a photon.

A surface plasmon (SP) is the collective electrostatic excitation of 
free electrons near a plasma dielectric surface. The importance of surface plasmons
at a plasma boundary was widely recognized following the pioneering
work of Ritchie \cite{r57}, and Trivelpiece and Gould \cite{tg59}. 
Their investigations were restricted to the propagation of electrostatic surface waves 
on a cold plasma half-space. The effect of a finite plasma temperature
has been studied later by Ritchie \cite{r63} using a hydrodynamical model,
and by Guernsey \cite{g69} using a more complete kinetic approach.
The frequency $\omega_{sp}$ of a surface plasmon on a flat surface of
a semi-infinite plasma can be easily determined from the frequency of 
the bulk plasma oscilation, as it corresponds to $\Re \, \epsilon (\omega_{sp}) 
= -\epsilon_d$, where $\epsilon$ and $\epsilon_d$ are dielectric
functions of the plasma and of the adjacent dielectric medium. For a cold 
plasma in contact with vacuum, one obtains $\omega_{sp} = \omega_{pe}/\sqrt{2}$.

The nonlinear propagation of the surface wave on a cold plasma half-space has
also been investigated \cite{yu78, sy02, sy03}. The latter reprted either the existence
of SP solitons, governed by a nonlinear Schr\"odinger equation \cite{yu78, sy03},
or the production and maintainance of a dense surface-wave driven plasma using a powerful
oscillator at twice the surface-wave frequency of the plasma \cite{sy02}.

\begin{figure}[h] \centering
    \includegraphics[width=60mm, height=50mm]{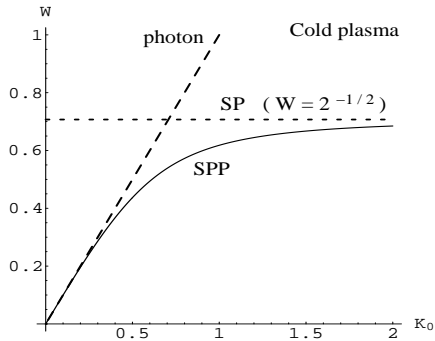}
\caption{Cold plasma half-space: dispersion curves for the 
surface plasmon-polariton (SPP) with a solid line and given by Eq. (\ref{e1}), 
the surface plasmon (SP) with long dashed line ($\omega = \omega_{pe} / \sqrt{2}$),
and the photon with short dashed line ($\omega = k_y c$). 
The coordinates are scaled as $W = \omega /\omega_{pe}$, and $K_0 = kc /\omega_{pe}$} \label{fig1} 
\end{figure}

Vedenov \cite{v65} provided a general description without limiting to 
electrostatic approximation, and showed that in a cold plasma
half-space, SPPs exist with frequencies ranging from 
$\omega_{spp} \to \omega_{sp} = \omega_{pe}/\sqrt{2}$ (in the electrostatic limit) 
down to $\omega_{spp} = 0$, as is shown in Fig. \ref{fig1}. 
Kaw and McBride \cite{km70} extended the plasmons analytical investigations
by including the effects of density gradient and the finite 
plasma temperature on the SPPs. 

We remember here the dispersion relation of the SPP on a cold plasma half-space
\be
k_x^{\rm vacuum} = -{k_x^{\rm plasma} \over \epsilon} 
\; \to \; \left(k_y^2 -{\omega^2 \over c^2}\right)^{1/2} = -{1 \over \epsilon} 
\, \left(k_y^2 -{\omega^2 \over c^2}\,\epsilon \right)^{1/2},  \label{e1}
\ee
where plasma permittivity $\epsilon = 1 - \omega_{pe}^2/\omega^2 <0$
must be negative (the plasma must have an overcritical density), so that 
the surface plasma wave propagates along the $y$-axis, i.e., $k_y^2 >0$, and decays
(is evanescent) in plasma, i.e. $k_x^2 < 0$.
Thus, the excitation of the surface plasmons is directly related to the problem of
the anomalous transparency of light in materials with negative
permittivities. In metallic nano-structures, the electron plasma is an
overcritical density plasma with a negative electric permittivity ($\epsilon <0$).
Media with negative electric permittivity and magnetic permeability
(the so-called left-handed materials - LHM) posses the possibility of 
creating an ideal lens with subwavelength resolution \cite{p00}. 
Transmission through such a superlens is essentially based on 
the excitation of surface plasmons \cite{p00} and the amplification of the 
evanescent waves due to their interference \cite{fv06}.

In metallic nano-structures no underlying ionic 
lattice exists, so that the electron dynamics is governed by 
the plasma effects \cite{sb95}. For such metallic plasmas, the quantum statistical effects 
should be included as long the electron Fermi temperature
is much higher than the room temperature. 
Moreover, the charge carriers density in miniaturized semiconductor 
devices is much lower than in metals, but their de Broglie wavelength can
be comparable to the spatial variation of the doping profile, 
and the quantum electron tunneling effect can no longer be ignored \cite{yn89}.

For a dense quantum plasma in metals or in intense laser produced solid density 
plasmas, the investigations included plasma echoes \cite{mf96} and
the quantum plasma modes and instabilities \cite{hm00, hg03,mk05,gh05}.
Major influence of the quantum effects have been proved for dense and 
dusty plasmas encountered in some representative astrophysical systems \cite{h05, sa05, as06, as07}.
It has been also demonstrated \cite{r07} that the bulk 
(high frequency) electrostatic oscillations can propagate in an 
underdense quantum plasma due to the quantum mechanical effects 
(the Bohm potential). In addition, the electromagnetic surface waves arising
in a plasma layer confined by a strong external magnetic field can be 
strongly affected by the quantum effects due to the magnetic 
inhomogeneity \cite{sr99}. 

\begin{figure}[h] \centering
    \includegraphics[width=60mm, height=50mm]{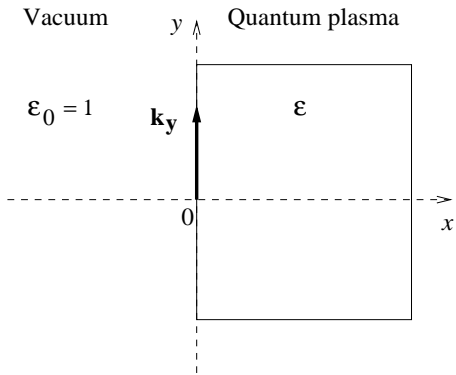}
\caption{Schematic diagram of a quantum plasma half-space: the interface
vacuum-plasma is located at the plane $x=0$, and the 
surface wave is propagating along the interface, ${\bf k} \parallel y$-axis.} \label{fig2} 
\end{figure}

In this Brief Communication we present the dispersion relation for 
surface plasmons that can exist on a dense quantum plasma half-space 
(see in Fig. \ref{fig2}).
The quantum model of the SPP mode is developed in Section II, 
where we derive its dispersion  
relation by employing the full set of the quantum hydrodynamical 
and Maxwell equations (see the formalism used by Kaw 
and McBride \cite{km70} for classical thermal plasma). 
Both the quantum statistical pressure and the
quantum electron tunneling effects are included here, 
while imobile ions are assumed to form the neutralizing background.
Further more, in Section III, we focus our attention on the electrostatic limit 
and present the dispersion relation for the electrostatic surface wave 
(SP) on a dense quantum plasma half-space. The SP mode is numerically 
evaluated for the specific case of a gold metallic plasma at room
temperature. The results are summarized in the Section IV.

\section{Quantum model of SPP}

We consider an homogeneous quantum plasma half-space as shown in Fig. \ref{fig2}.
The dynamics of the electrons are governed by the continuity equation

\be
{\partial n \over \partial t} + n_0 \nabla \cdot {\bf v}=0, \label{e2}
\ee
and the momentum equation

\be
{\partial {\bf v} \over \partial t} =- \,{e \over m}\,{\bf E} -{2 T_{Fe} \over m n_0} 
\nabla n + {\hbar^2 \over 4 m^2 n_0} \nabla \nabla^2 n, \label{e3}
\ee
where we have included both the quantum statistical effects involving the electron Fermi temperature, $T_{Fe} = m v_{Fe}^2 /2$, and the quantum diffraction effect contained in the $\hbar$-dependent term (sometimes called the Bohm potential).  
Here $n$ ($\ll n_0$) is a small electron density perturbation in the
equilibrium number density $n_0$, $v$ is the electron fluid velocity,
$e$ is the magnitude the electron charge, $m$ is the electron mass, and 
$\hbar$ is the Plank constant divided by $2\pi$. The electromagnetic 
fields are given by the Maxwell-Poisson equations
\be
\nabla \times {\bf E} = - {1\over c} \, {\partial {\bf B} \over \partial t}, \label{e4}
\ee
\be
\nabla \times {\bf B}= -{4 \pi n_0 e \over c} {\bf v} + {1\over c} \, {\partial {\bf E} \over \partial t}, \label{e5}
\ee
\be
\nabla \cdot {\bf B} = 0, \label{e6}
\ee
\be
\nabla \cdot {\bf E} = -4 \pi e n, \label{e7}
\ee
The wave equation for the electron density perturbation is obtained from (\ref{e2}),
(\ref{e3}) and (\ref{e}) by assuming that all physical quantities (electron 
density perturbation and electromagnetic field perturbation) vary as 
$\Psi(x) \exp (ik_y y - i \omega t)$,
\be
{\partial^2 n \over \partial x^2} - \gamma_p^2 n =0, \label{e8}
\ee
where $k_y$ is the component of the wave vector along the $y$-axis (which
is the direction of the dense plasma - vacuum interface), $\omega$ is the frequency, 
and 
\be
\gamma_p = \left(k_y^2 + {\omega_{pe}^2 - \omega^2 \over v_{Fe}^2 + 
{\hbar^2 k_y^2 \over 4 m^2 }} \right)^{1/2}. \label{e9}
\ee
The very slow nonlocal variations are neglected, $\partial^4 /\partial x^4 \ll 
\partial^2/\partial x^2 \ll k_y^2$. The wave equation for the magnetic field of the 
SPP is
\be
{\partial^2 {\bf B} \over \partial x^2} - \alpha_p^2 {\bf B} =0, \label{e10}
\ee
where
\be
\alpha_p = \left(k_y^2 + {\omega_{pe}^2 - \omega^2 \over c^2} \right)^{1/2}. \label{e11}
\ee

Equations (\ref{e8}) and (\ref{e10}) have the solutions, respectively,
\be
n = A \exp (- \gamma_p x),\;\;\; x>0, \label{e12}
\ee
\be
{\bf B} = {\bf F_1} \exp (\alpha_v x),\;\;\; x<0,  \label{e13}
\ee
\be
{\bf B} = {\bf F_2} \exp (- \alpha_p x),\;\;\; x>0,  \label{e14}
\ee
where for vacuum, $\alpha_v = (k_y^2 - \omega^2 /c^2)^{1/2}$, 
anf $A$ and ${\bf F_{1,2}}$ are constants. Using again the
Maxwell equations, we find for the electric field of the SPP
\be
{\bf E} = {\bf D_1} \exp (\alpha_v x),\;\;\; x<0,  \label{e15}
\ee
$
{\bf E} = {\bf D_2} \exp (- \alpha_p x) 
$
\be 
- 4 \pi eA \left[v_{Fe}^2 - 
{\hbar^2 (\gamma_p^2 -k_y^2) \over 4 m^2 } \right] \, 
{\exp (- \gamma_p x) \over \omega_{pe}^2- \omega^2} 
\,[-\gamma_p \hat{\bf x} +i k_y \hat{\bf y}],\;\;\; x>0,  \label{e16}
\ee
where ${\bf D}_1$ and ${\bf D}_2$ are constants. For the interface plane located at $x=0$, 
as shown in Fig. \ref{fig2}, we keep only that part of 
solutions that decay away from the interface in both regions.

Imposing the continuity condition at the boundary $x=0$
for the electromagnetic field components contained into the
interface plane, together with the boundary condition 
$v_{x} = 0$ at $x = 0$ for the plasma particles, we obtain the 
following dispersion relation
\be 
\omega_{pe}^2 k_y^2 = \gamma_p [\omega^2 \alpha_p - 
(\omega_{pe}^2-\omega^2) \alpha_v]. \label{e17}
\ee
To our knowledge, this is the first time derived 
dispersion relation of SPP mode on a quantum plasma
half-space, including quantum effects due to the
Fermi electron temperature and the Bohm potential.

If we assume over-critical density plasmas, so that
$k_y^2 v_{Fe}^2 + \hbar^2 k_y^4 / 4 m^2  \ll | \omega_{pe}^2 - \omega^2|$,
then (\ref{e17}) yields

\be
{\omega_{pe}^2 - 2 \omega^2 \over \omega_{pe}^2 - \omega^2}
= {\omega^2 \over k_y^2 c^2} - {2 (k_y^2 - \omega^2 /c^2)^{1/2} \over 
\sqrt{\omega_{pe}^2 -\omega^2}} \left[v_{Fe}^2 + 
{\hbar^2 k_y^2 \over 4 m^2 } \right]^{1/2}. \label{e18}
\ee
Without the quantum effects, i.e., $v_{Fe}^2 + \hbar^2 k_y^2 / 4 m^2  \to 0$,
we recover from (\ref{e18}) the well-known dispersion relation, 
given in Section I, for the SPP on a cold plasma
half-space.

\section{Surface electrostatic waves}
 
\begin{figure}[h] \centering
    \includegraphics[width=60mm, height=50mm]{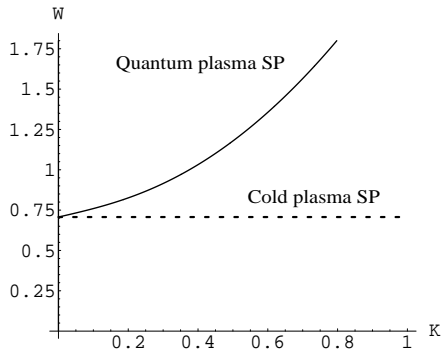}
\caption{{\bf Solid line}: Surface electrostatic modes on a quantum gold plasma half-space 
(contributions of quantum statistical effect and Bohm potential are included); 
{\bf Dashed line}: Surface electrostatic modes on a cold plasma half-space. Here
$W = \omega/ \omega_{pe}$ and $K = k_y v_{Fe} /\omega_{pe}$.} \label{fig3} 
\end{figure}

At room temperature and standard metallic densities,
the transverse electromagnetic component of SPP is 
unaffected by the quantum correction, (i.e. plotting Eq. (\ref{e18})
we find the same dispersion curve as in Fig. \ref{fig1}),
and therefore we restrict our attention to the electrostatic surface waves only.
The electrostatic limit of (\ref{e18}) is taken by letting 
$c \to \infty$, whence we find that
\be
\omega = {\omega_{pe} \over \sqrt{2}} \left(1 + {k_y v_{Fe} \over \sqrt{2} \omega_{pe}}
\sqrt{1 + {\hbar^2 k_y^2 \over 4 m^2 v_{Fe}^2 }} \right). \label{e19}
\ee
Introducing the normalized quantities 
$W = \omega / \omega_{pe}$, $K = k_y v_{Fe}/ \omega_{pe}$,
and $H = \hbar \omega_{pe} / (2 m v_{Fe}^2)$,  
we rewrite (\ref{e19}) as
\be
\Omega = {1 \over \sqrt{2}} \left(1 + {K \over \sqrt{2}}
\sqrt{1 + H^2 K^2 } \right), \label{e20}
\ee
which is plotted in Fig. \ref{fig3} for typical parameters
of the gold metallic plasma at room temperature \cite{m05}: 
$n_0 = 5.9 \times 10^{22}$ cm$^{-3}$, 
$\omega_{pe} = 1.37 \times 10^{16}$ s$^{-1}$, $v_{Fe} = 1.4 \times 10^8$ cm/s, 
and therefore $E_{Fe} = 5.53$ eV $\gg E_T = 0.026$ eV.
The quantum effects are visible at smaller scales 
of the order of $v_{Fe} /\omega_{pe} = 10^{-8}$ cm $\ll c/\omega_{pe} 
\simeq 2.4 \times 10^{-6}$ cm.

The 'plasmonic' coupling parameter, $H= \hbar \omega_{pe} / (2 m v_{Fe}^2)$, 
used in (\ref{e20}), describes the ratio of the plasmonic energy density 
to the electron Fermi energy density. 
Ignoring the effects caused by the Bohm potential, $H \to 0$,
we recover from (\ref{e19}) or (\ref{e20}) the equation derived by Ritchie for
the surface electrostatic waves on a thermal plasma half-space.
However, for typical parameters of a gold metallic plasma at room temperature the
Bohm potential term cannot be neglected.

\section{Conclusions}

We have investigated the surface waves on a quantum plasma half-space including the effects
of a quantum statistical Fermi electron temperature and the quantum electron tunneling. The quantum hydrodynamical model used here includes the full set of Maxwell equations, which allowed us to derive the dispersion relation for the SPP mode.
In the electrostatic limit, we derived the dispersion relation for electrostatic surface SP mode. 
The specific case of gold metallic plasma is considered to show that, at room temperature, the
electrostatic surface waves are significantly affected by the quantum effects.
In this case we plotted the dispersion curve, which shows that the quantum effects facilitate the propagation
of the electrostatic surface wave. The present result is of practical utility to 
understanding the dispersion properties of electron plasma oscillations at the 
dense plasma - vacuum dielectric interface.

\begin{acknowledgments}
This work was partially supported by the NSERC Canada and
by the Deutsche Forschungsgemeinschaft through the Sonderforschungsbereich 591.
\end{acknowledgments}

\end{document}